\documentclass{emulateapj}

\shorttitle{Space Velocities of Globular Clusters}

\begin{document}

\title{Space Velocities of Southern Globular Clusters. VI. Nine Clusters in 
the Inner Milky Way}

\author{Dana I. Casetti-Dinescu\altaffilmark{1,2}, 
Terrence M. Girard\altaffilmark{1}, 
Vladimir I. Korchagin\altaffilmark{4,1},
William F. van Altena\altaffilmark{1} and 
Carlos E. L\'{o}pez\altaffilmark{3}}

\altaffiltext{1}{Astronomy Department, Yale University, P.O. Box 208101,
New Haven, CT 06520-8101, USA}
\altaffiltext{2}{Astronomical Institute of the Romanian Academy, Str.
Cutitul de Argint 5, RO-75212, Bucharest 28, Romania}
\altaffiltext{3}{Yale San Juan Observatory, Universidad Nacional de San Juan,
Avenida Benavidez 8175 Oeste, 5407 Marquesado, San Juan, Argentina}
\altaffiltext{4}{Institute of Physics, Rostov University, Rostov-on-Don 344090, Russia}
\email{dana.casetti@yale.edu}

\begin{abstract}
We have measured the absolute proper motions of nine low-latitude, inner Galaxy 
globular clusters, namely NGC 6273 (M 19), NGC 6284, NGC 6287, NGC 6293, NGC 6333 (M 9), NGC 6342, 
NGC 6356, NGC 6388 and NGC 6441. These are the first determinations ever made for these clusters.
The proper motions are on the International Celestial Reference System via $Hipparcos$. 
The proper-motion errors range between 0.4 and 0.9 mas yr$^{-1}$ and are dominated by 
the number of measurable cluster members in these regions which are 
very crowded by the bulge/bar and the thick disk.
This sample contains five metal poor ([Fe/H] $< -1.0$) and four 
metal rich ([Fe/H] $\ge -1.0$) clusters; seven clusters are located within $\sim 4$ kpc
from the Galactic center, while the remaining two, namely NGC 6356 and NGC 6284 
are in the background of the bulge at
$\sim 7.5$ kpc from the Galactic center. 

By combining proper motions with radial velocities and distances from the literature we derive
3D velocities. In a number of cases, distance uncertainties make the kinematical classification 
ambiguous. For the metal-poor group of clusters, we obtain
that three clusters, namely NGC 6273, NGC 6287 and
NGC 6293 are members of a kinematically hot system, the inner halo. 
As for the remaining two metal poor 
clusters,
NGC 6284 is located at $\sim 7.5$ kpc from the Galactic center and  
kinematically belongs to the thick disk, while
NGC 6333, located in the inner $\sim 2 $ kpc, has an uncertain membership
(between halo and thick disk) due to the distance uncertainty.

Within the metal rich group of clusters, NGC 6356 and NGC 6342 have velocities 
compatible with membership
in the thick disk; however, velocity uncertainties do not allow us to rule out
their belonging to a hotter system. NGC 6342 is within $\sim 2$ kpc
from the Galactic center, and thus it may belong to the bulge. NGC 6356 is
at $\sim 7.5$ kpc from the Galactic center and thus its metallicity, kinematics and location
argue together in favor of thick-disk membership.

The remaining  two metal rich clusters, NGC 6388 and NGC 6441 have velocities
incompatible with membership in the thick disk or the bar of the Milky Way. They can be thought of as
members of a kinematically hot system in the inner Galaxy. Curiously, both clusters have 
similar velocity components. Together with their similar Galactic location,
and peculiar but similar stellar-population characteristics,
these two clusters may share a common origin. 
Their velocities are also very low indicating that the two
clusters are now at apocenter, i.e., they will not leave the inner $\sim 4$ kpc of the Galaxy.

\end{abstract}


\section{Introduction}

Substantial progress has been made in recent years in our understanding
of the bulge region of our Galaxy thanks to large-area near infrared photometric 
surveys as well as spectroscopic surveys that have provided radial velocities and
detailed chemical abundances for large samples of stars (Zoccali 2009).
Yet, due to the complexity of this region, we are now confronted with 
many details,
but not a comprehensive, self-consistent picture. Recent kinematical
evidence by Howard et al. (2009) combined with information from 
COBE/DIRBE observations, as well as starcounts, show the bulge to have a boxy morphology
with cylindrical rotation (see review by Freeman 2008) consistent with
its formation via a disk instability. 

Alternatively, the bulge 
could form via early mergers and end up as a merger remnant with a spherical distribution
and be rather hot kinematically. Indeed this is what is seen when one analyzes
radial velocities of
massive ($M_V < -7.9$), metal rich ([Fe/H] $> -0.8$) clusters in the inner 
($r_{GC} \le 4$ kpc) Galaxy (Burkert \& Smith 1997, C\^{o}t\'{e} 1999). These studies
show however that low-mass, metal-rich clusters have kinematics consistent with bar membership.
Since the kinematical properties are derived only from radial velocities, the overall geometry and
other assumptions, the distinction between these groups is not clear-cut.
Likely thick disk cluster members confuse the picture as well.
To the rescue come tangential velocities which enable us to build a clearer picture 
of the inner Galaxy as inferred from its globular clusters. It is important to see whether 
globular clusters resemble the field stellar population of the bulge and to what degree, 
if we wish to have a comprehensive formation picture of the bulge.

Here, we continue our previous work on absolute proper motions of globular clusters
based on Southern Proper Motion Program data. 
We describe the measurements in Section 2, the derivation of the absolute proper motions in 
Section 3, and that of the velocities in Section 4. We discuss our results in Section 5.

\section{Measurements}

We continue our program to measure absolute proper motions of globular clusters
as a part of the overall Southern Proper Motion Program (SPM) described in a series of
papers: Platais et al. (1998), Girard et al. (1998), 
Girard et al. (2004, 2010). 

The cluster part of the SPM provided measurements
for 25 clusters reported in a series of papers: Dinescu et al. (1997, 1999ab, 2003) and 
Casetti-Dinescu et al. (2007), named Papers I through V, respectively.
Together with the sample presented here, the SPM has measured
34 clusters, which represents $55\%$ of the total number of clusters 
with such measurements\footnote{An updated version of the list of all globular clusters with absolute 
proper motions can be found at www.astro.yale.edu/dana/gc.html}.

The first-epoch SPM survey was taken from 1965 to 1979, and 
is entirely photographic. The SPM second-epoch survey is approximately one third photographic
(taken from 1988 to 1998) and two thirds CCD-based (taken from 2003 through 2008).
The main survey CCD data started with a two-fold overlap of the frames (see below).
Later, single coverage was used, when it was found that a single CCD exposure was 
already astrometrically superior to the first-epoch plate material. For the clusters, apart from
the main survey, we take additional frames centered on each cluster. In Table 1 we
list the SPM field centers, and the photographic and CCD material used in the reductions,
together with the epochs of the observations.

In three of our four fields presented here, the second-epoch positions are CCD-based. In the remaining
field (SPM \#436, cluster NGC 6441), the second-epoch positions are predominantly photographic, 
supplemented by six CCD exposures taken on the cluster itself.
The three fields with no second-epoch photographic material lacked a small portion 
of CCD coverage from the main SPM survey. 
These areas consist of a declination band
for field $\# 578$ amounting to $23\%$ of the entire area, two corners of field $\# 307$
amounting to $4\%$ of the entire area, and about $1\%$ for $\# 651$ around the edges of the field
(see Table 1).
For these limited areas, the second-epoch positions are taken from $Hipparcos$, 
Tycho-2 (Perryman et al. 1997), UCAC2 (Zacharias et al. 2000) and 2MASS 
point source catalog (Cutri et al. 2003). Given the 
construction of the input list of objects to be measured, these positions are 
predominantly from 2MASS.

The program clusters are at low Galactic latitude and toward the bulge direction. Since 
extinction is high in these directions, we used $Hipparcos$ stars to tie to an inertial 
reference system rather than extragalactic objects. Thus the proper motions are on the 
International Celestial Reference System via $Hipparcos$. 
Proper-motion units throughout the paper are mas yr$^{-1}$.

\subsection{Photographic Measurements}

The SPM plates were taken with the 51-cm double astrograph at Cesco Observatory in two passbands:
blue (103a-O, no filter) and visual (103a-G + OG515 filter). The plate scale is 55.1''/mm, 
and each field covers $6.3\arcdeg \times 6.3\arcdeg$. The plates contain two exposures: 
one of 2 hours that reaches $V\sim 18$ and an offset of 2 minutes. During both exposures, 
an objective grating is used which produces a series of diffraction images on 
either side of the central, zero-order images. 
The multiple sets of images for bright stars (for $V \ge 14$, there are only long exposure,
zero-order images) allow us to detect and model magnitude-dependent systematics that 
affect positions and consequently proper motions.
The method has been described and thoroughly tested by Girard et al. (1998), and it shows
that $Hipparcos$ stars with $V \sim 9$ and faint cluster stars ($V \ge 14$) can be placed on
a system that is largely free of systematics. For very bright magnitudes ($V < 8-9$, 
the corrections are large and likewise their uncertainties; thus it is possible 
that residual magnitude equation is left at the very bright end, and for this reason, the very
bright $Hipparcos$ stars are not used. Since the magnitude correction is made independently for each plate,
and there are multiple same-epoch plates (from 2 to 4) in each field, residual magnitude equation errors 
will show up as scatter in the multiple-plate data and contribute to the formal proper-motion uncertainties
derived from positional scatter (see Section 3).

The plates are measured with the Yale PDS microdensitometer in an object-by-object mode with
a pixel size of 12.7 $\mu$m. For each SPM field, we measure a preselected set of
stars (see also Papers IV and V). This set consists of all $Hipparcos$ stars 
(typically 100 per field), a set of 
$\sim 1800$ Tycho-2 stars (Perryman et al. 1997), another set of $\sim 1000$ bright 
($R_{UCAC2} < 13.5$) UCAC2 stars (Zacharias et al. 2000), 
reference stars and cluster-region stars. Reference and cluster stars are 
selected from UCAC2  and 2MASS (Cutri et al. 2003) catalogs.  Cluster-region stars 
are selected to reside within a $7\arcmin$ 
radius from the cluster center. Reference stars consist of two subsets: one set uniformly 
distributed over the entire field, the other as a ring around each cluster with radius  
$7\arcmin \le r \le 11\arcmin$. Within the ring and cluster area, all objects found in 
UCAC2 or 2MASS are selected for the input list.
 For the uniformly distributed reference stars over the 
entire field, we select a fraction of UCAC2 ans 2MASS stars imposed by the time limitation of the PDS scans.
For instance, for field $\# 578$ where there are four clusters, we had a total input number of 
$\sim 3900$ stars for the uniformly distributed reference stars, $\sim 13600$ for the ring stars, and
$\sim 11500$ stars for cluster-region stars. For field $\#307$, where there is one cluster,
we had a total input number of 
$\sim 12700$ stars for the randomly distributed reference stars, $\sim 7100$ for the ring stars, 
and $\sim 3800$ for the cluster-region stars.

The bright stars in the input catalog assure an appropriate magnitude range of various diffraction 
orders with which to model magnitude-dependent systematics. Only $Hipparcos$ stars are 
used to determine the correction to absolute proper motions.
The reference stars (ring and entire field) are used to map plate positions into one another 
as well as CCD positions into the photographic positions. 

For one field, $\# 436$, we had a slightly different input catalog, as this field was measured
early on, together with our Paper IV clusters, since both epoch plates existed.
However, at that time, we realized that although we can obtain a reliable correction for the
absolute proper motion for this field, the motion of cluster NGC 6441 was unreliable. This 
cluster is more distant, and the time baseline for the plates is only 22 years (Tab. 1). So, we 
took several CCD frames centered on the cluster in 2007 to improve the mean motion of the cluster.
The strategy used to make the input catalog was similar to the one presented above; however instead of 
UCAC2 and 2MASS, we used the Guide Star Catalog (GSC) 1.1 (Lasker et al. 1990) and the 
USNO-A2.0 catalog (Monet et al. 1998). For cluster stars, we derived a catalog based on 
a coarse scan centered on the cluster of the deepest SPM plate. 

During these long PDS scans (up to 24 hours) we monitor a set of 8 to 10 stars that are used to correct for
thermal drift. Positions, instrumental magnitudes and other image parameters are
derived with the Yale centering routines that fit a two-dimensional Gaussian function to 
the image profile (Lee \& van Altena 1983).

\subsection{CCD Measurements}

The CCD system mounted on the double astrograph consists of two main cameras: a 4K x 4K 
PixelVision (PV) camera (15 micron pixel) in the focal plane of the yellow lens and an Apogee 1K x 1K
(24 micron pixel) camera in the focal plane of the blue lens. Later, the Apogee camera
was replaced with an Apogee Alta 2K x 2K (12 micron pixel) camera. The PV's field of view is
$0.93\arcdeg \times 0.93\arcdeg$ with a scale of 0.83''/pix. The Apogee Alta camera covers
a $0.38\arcdeg \times 0.38\arcdeg$ area with a scale of 0.74''/pix. The PV data were used for 
both astrometry and photometry, while the Apogee data were used only for photometry.
The observations are taken with the diffraction grating oriented at $45\arcdeg$, thus 
ensuring a dynamical range of 10 mag (i.e., 6 from the CCD plus 4 from the grating).
Exposure times are 120 sec, reaching the same depth as the first-epoch plates.
Between 3 and 6 frames are taken on each cluster in addition to the survey CCD frames taken 
for each SPM field (Tab. 1). These have exposures of 30, 60 and 120 sec.
The astrometric reduction of the CCD data is thoroughly described in Paper V; here we 
briefly mention only the steps applied.

The CCD data are corrected for bias, dark (only in the case of the Apogee frames is it significant)
and flat fielding. SExtractor (Bertin \& Arnouts 1996) is used for object detections, 
aperture photometry and preliminary x,y centers. Final x,y centers are derived by fitting
two-dimensional elliptical Gaussian functions to image intensities. These image centers have 
a positional precision of $\sim 20$ mas per single image, for $V=7-15$.
Two other corrections of the CCD positions are applied per frame. The first is a correction for
the optical field angle distortion (OFAD). This correction is determined from 
stacked residuals into UCAC2, and is applied as a mask to the CCD data (see Paper V). 
The second correction refers to the positions of different grating-order images that must be placed
on a common system with those of the central-order images.  We found offsets between the 
average position of symmetric image orders and the position of the central order, but these offsets 
are not related to the magnitude of the object (as is the case for photographic plates). 
Therefore these offsets were corrected as such, a simple offset between different  image 
order systems (see also Paper V).

The next step is to link all of the CCD frames for a given SPM field from the main survey 
to produce a $6\arcdeg\times6\arcdeg$ CCD pseudo-plate. This is different from our procedure 
adopted in Paper V, where
each individual CCD frame was mapped into a master first-epoch plate.
The linking of the CCD frames was developed for the construction of the SPM4 catalog and is
described in Girard et al. (2010).  
In brief, the mask-corrected
CCD x,y positions of each frame are transformed
onto the system of the UCAC2 to facilitate pasting together the roughly 50 
to 100 frames (depending on whether the field had two-fold or single coverage) 
that comprise a $6\arcdeg\times6\arcdeg$ SPM field.  An iterative overlap method is 
employed to
perform this task using Tycho2 stars as an external reference system to
ensure that systematics from the individual frame overlaps do not
accumulate.  In this manner, an artificial "pseudo-plate" is built up from
CCD frames.  The CCD frames that contribute to any one field's pseudo-plate
may actually span two to three years' range in epoch whereas an actual plate
has a single epoch.  For this reason,
once preliminary proper motions are determined from a second-epoch 
pseudo-plate, an improved pseudo-plate is constructed, one that incorporates
the preliminary proper motions prior to pasting the frames together in a
new overlap solution. 
This single-epoch pseudo-plate can then be treated the same as a real 
second-epoch plate in the reduction process to yield final proper-motion
determinations.

The CCD $BV$ photometry is derived from aperture photometry calibrated into 
Tycho2 (corrected to the Johnson system) and several other external catalogs
such as the Guide Star Photometric Catalog II (Bucciarelli et al. 2001).
When no CCD photometry is available,
we use photographic photometry from the first-epoch plates calibrated on Tycho2.
For our purposes in this paper, we use however 2MASS photometry.

\section{Proper Motions}

To obtain proper motions we use the central-plate overlap method described for instance by
Girard et al. (1989). Thus, we transform all measurements --- CCD pseudo-plate, on-cluster CCD frames,
photographic positions of various image grating-order systems --- into one master photographic
plate list which is chosen to be a visual plate with the central-order image. This transformation
is a polynomial function of x,y coordinates of up to 5th order, and it uses stars with 
magnitudes between $V = 10$ and 17 distributed uniformly over the field and with a higher density
in a ring around each cluster. The number of reference stars for different epoch
measurements ranges between 500 and 6000 stars,
with a mean average magnitude roughly between $V = 14$ and 16.
A linear least-squares fit of positions as a function of time gives the proper motion of each object.
Measurements that differ by more than $0.2\arcsec$ from the best fit line are rejected as outliers.
The formal proper-motion error is given by the scatter around the best-fit line.

For field $\# 436$ we have applied a different procedure. 
The photographic based motions are
derived from each plate pair: i.e. blue second-epoch into blue first-epoch,
and likewise for the visual plate pair. This ensured proper motions for stars over the entire field,
and provided the correction to absolute proper motions via $Hipparcos$ stars. For the cluster,
we have applied a central-plate overlap method to include the CCD frames taken on-cluster.
We have selected reference stars in the same magnitude range for the cluster solution 
as was used in the plate-pair, entire-field solution. In this way we can apply 
the correction to absolute as an offset, assuming that the reference stars belong to 
a kinematically similar population.

\subsection{Correction to Absolute Proper Motions} 

The correction to absolute proper motion is determined directly from the difference between
the $Hipparcos$ proper motion and our relative proper motion. These differences also serve 
to check for various systematics as a function of magnitude and position on the plate.
In Figure 1 we show the proper-motion difference in each celestial coordinate
$\Delta \mu = (\mu_{Hipparcos} - \mu_{SPM})$ as a function of $V$ magnitude for the three fields
that had only CCD data as second epoch. In what follows we use the notation 
$\mu_{\alpha}^* = \mu_{\alpha} cos \delta$.
From Fig. 1 it can be seen that at the bright end, the magnitude equation is not entirely corrected.
For this reason, we eliminate the bright $Hipparcos$ stars; the bright magnitude limit is 
indicated with a vertical line. Formal SPM errors in the proper-motion are between
2.1 and 2.6 mas yr$^{-1}$ for these stars, and are represented with a gray symbol in Fig. 1.

In Figure 2 we show the proper-motion differences in both coordinates in a vector-point diagram
(VPD) (left panels), and as a function of position on the plate (middle and right panels).
Cluster locations are marked with a gray vertical line. In general, there are no significant trends
with position, except toward the edges of the field in some cases.
The final correction to absolute proper motion --- also referred to as zero point correction ---
per field is determined using 
probability plots (Hamaker 1978) trimmed at $10\%$ at each side. These values are listed in
Table 2, and represented with a cross in Fig. 2.
In Table 2 we also include a column with the clusters in each field.
 
For field $\#436$, we show the values of the correction to absolute 
proper motion obtained from two plate pairs and from various image orders in
Figure 3.
There are three image-order solutions for the long exposure and three for the
short exposure, thus six in all per plate pair.
Open symbols show the visual plate solutions, filled the blue plate solutions.
Here too we used only $Hipparcos$ stars that are not affected by residual 
magnitude equation, i.e. $V > 7.0$. The mean values for each solution are 
obtained using probability plots trimmed at $10\%$ on each side.
The one deviant point is the central order 
of the visual plate pair, and we exclude it from the final mean.
The final mean value for the field is the error-weighted mean of 
the eleven determinations, and it is represented with a red symbol in Fig. 3.
Its error is given by the scatter of the individual solutions.

\subsection{Cluster Proper Motions}

The most challenging part of the measurements presented here is the 
appropriate choice of cluster members. The clusters 
lie in regions strongly contaminated by the Milky Way bulge, bar and
disk.  In Figure 4 we show the location of the clusters in Galactic
coordinates, together with the 2.2 $\mu$m surface brightness
data from Launhardt et al. (2002) presented in their Figure 5.
Note the particular location of NGC 6441 which is discussed
further, later on.  It presented a 
particularly difficult case of contamination. 

One way to minimize this contamination is to work with a small area
centered on the cluster.
If however we restrict our candidate cluster stars to a small area centered on
the cluster (of radius $ < 4\arcmin$) we face the problem of crowding 
and thus poor astrometric precision. Within $1\arcmin-2\arcmin$ of a cluster center our 
photographic plates are essentially unusable, and thus we are left with 
rather few well-measured cluster stars, especially for less massive clusters.
To overcome this problem, we first work only with visual plates which have
a tighter PSF than the blue plates, and second we select candidate 
cluster members from 2MASS color-magnitude diagrams (CMDs).
For five clusters we use as a guideline the ridge line of the giant branch
determined in the same passbands as 2MASS by Valenti et al. (2007); in some
cases we use their photometry as well as a guide to select 2MASS-based 
members.  The Valenti et al. (2007)\footnote{the database is at 
www.bo.astro.it/~GC/ir-archive} study 
provides accurate infrared photometry for bulge-region
 clusters, and it
therefore has better photometry than 2MASS in the cluster region. However
we could not use only this photometry since it covers a
$4\arcmin \times 4\arcmin$ box centered on the cluster. Our 
cluster area is within $7\arcmin$ of the cluster center, except NGC 6441
where we used a $23\arcmin \times 23\arcmin$ box centered on the cluster.
These choices have to do with the construction of the input catalogs (see
Section 2.1).
 
In Figure 5 we show the CMDs for the ring region (first column from the left)
and for the cluster region (second column). 
The radius range of these regions is specified in each panel.  
Each row represents one cluster labeled in the panels of the second column.
Here we also indicate the cluster distance, metallicity and $E_{B-V}$ reddening
as listed in the 2003 update of the Harris catalog 
(Harris 1996, hereafter H03). The CMDs show objects with measured proper motions 
(rather than the complete 2MASS catalog in the given region). In red we show our 
selection of red giant cluster candidates.
The ridge line from Valenti et al. (2007) is shown with a black line.
The cluster sequence is visible as an overdensity of stars in the 
cluster-region selected samples, compared to the ring region.
The third column of Fig. 5 shows the relative proper motion VPD
for all objects in the cluster-region sample (i.e., in the CMD from the 
second column). The fourth column shows the VPD of CMD-selected cluster 
candidates. The proper-motion marginal distribution in each coordinate is overplotted.
The black cross-hairs indicate the value of the adopted 
mean relative cluster motion.
Figure 6 is similar to Fig. 5, but for the next set of four clusters.
The faint limit of the samples is imposed by the morphology of the 
giant branch of the cluster in relation to the field population.
Clearly, at faint magnitudes there is more overlap with the field 
population than at bright magnitudes.
Also, we do not use stars fainter than $K = 14$ as these are at the faint
limit of the proper-motion data, where errors are large.

For these eight clusters, we have adopted two different ways to estimate 
the mean relative proper motion according to the size of the CMD-selected 
cluster candidates. First, we eliminate high proper-motion stars, 
i.e., with $|\mu| \ge 20$ mas yr$^{-1}$.
Then, for small samples we use probability plots trimmed at
$10\%$ on each side. This implies that in the CMD-selected sample 
only $80\%$ of the stars are cluster stars, and the rest
 are treated as contaminants. We have also estimated this contamination by using
the CMD from the ring area, scaled to the cluster-region area, and we
found that  $20\%$ contamination is a realistic value.
The results are insensitive to different trimming fractions (between 5 and 15\%).

For larger samples --- which are mainly those 
of rich, massive clusters --- we fit the sum of two Gaussian functions to the 
proper-motion distributions in each coordinate: one representing the field,
the other the cluster population (see e.g., Girard et al. 1989).
The error in the mean motion is determined from the width of the cluster 
distribution divided by the square root of the number of cluster stars, both
given by the fit. From the fits of each coordinate, we adopt the more
conservative error, i.e. the larger in size.

In Table 3 we list the mean relative proper motion obtained for each cluster,
together with the faint $K$-band magnitude limit of the CMD selected sample,
the number of stars in the sample and the method adopted to derive the mean.
Numerous tests have been done for each cluster with the CMD-selected sample
(i.e. fainter magnitude limits), radius-selected sample (i.e., smaller than 
$7\arcmin$), and 
with the means estimated using more severely trimmed probability plots, to
finally arrive at the solutions adopted in Table 3, which we regard as 
robust within the quoted errors.

A more difficult case is NGC 6441, that lies in a complex Galactic
region. In Figure 7 we show the data for NGC 6441. The top left panel
shows the CMD of a large ring 
region centered on the cluster ($10\arcmin < r < 20\arcmin$)
as given by the entire 2MASS catalog in this region. With a red line we show
the giant-branch ridge line determined by Valenti et al. (2007). 
This plot shows a rather well-defined giant branch which is
displaced from the ridge line defined by the cluster.
At a given magnitude, it is slightly redder than the cluster ridge line,
 and  extends to brighter magnitudes.
Since it is a rather narrow feature, we suspect that the majority of the
stars in this structure are at a given distance which is 
roughly similar to that of the cluster. 
Likewise, the shape of the giant branch argues for a metallicity 
somewhat similar to that of the cluster.
Since NGC 6441 is at a distance of 11.7 kpc from the Sun, it is 
unlikely that this giant branch belongs to the concentration of stars
at the Galactic center, or ``nominal'' bulge stars from a spherical distribution
with its peak at the Galactic center.
Examining the location of NGC 6441
with respect to the Galactic bar (with parameters as summarized by 
Gerhard 2002, see also Fig. 4 and Section 5),  
it appears that the cluster is located near the far 
end of the bar, with a geometry where the line of 
sight from the Sun toward the cluster crosses the tip of 
the far end of the bar. That this is the case, can also be gleaned from 
the work of Babusiaux \& Gilmore (2005), where they trace the bar out to 
longitudes of $\pm 9.7\arcdeg$ by studying red clump stars from a
near-infrared photometric study. Their field located at $l  = -5.7\arcdeg $
clearly detects the far side of the bar at a galactocentric radius of $\sim 2$ kpc,
or a distance from the Sun of 9.9 kpc. This field is near the location of NGC 6441
(see Tab. 4), only closer to the Galactic plane ($b \sim 0.2\arcdeg$).
Babusiaux \& Gilmore (2005) also note that their detection at $l= -9.7\arcdeg$
has a distance inconsistent with the continuation of the bar at these longitudes, 
and they attribute this detection and its distance to the signature of the end of the bar
which is circumscribed by an inner ring.

Therefore this well-defined giant branch is likely due to the 
far end of the bar, (or to a ring)
located in the foreground of the cluster (if we adopt the 
geometry from Babusiaux  \& Gilmore 2005).
Given that it lies under the ridge line of NGC 6441, it is
probably more metal rich than NGC 6441.

This structure makes our choice of cluster members
very difficult. In the top right panel of Fig. 6 we show
the CMD of a smaller area centered on the cluster for stars with 
proper motions. The giant branch seen in the previous plot is
present here too, lying on top of the cluster giant branch.
Because of this confusion we seek other criteria to 
select cluster members. Fortunately, NGC 6441 has been observed 
spectroscopically more recently by Gratton et al. (2007) where 
cluster giants are selected based on the CMD, radial velocity and distance from
cluster center criteria. They present a set of 25 likely members; of 
these, 16 stars have proper motions. These stars are shown with 
filled symbols in Fig. 7 top-right panel. The stars used for the
mean proper motion of the cluster are shown with red filled symbols.
The bottom-left panel of Fig. 7 shows the 
relative proper-motion VPD for stars in the cluster area
(those shown in the top right CMD). The bottom-right panel shows our 
proper motions for the cluster members from Gratton et al. (2007).
A proper-motion clump is apparent with a scatter comparable with the
formal errors. To determine the mean proper motion, we use stars 
with four or more
plate/CCD measurements, and eliminate stars with deviant proper
motions from the mean: the stars contributing to the mean are shown with red symbols. 
The dark circle marks the adopted mean which is a simple average of
the proper motions of seven stars. The error of the mean is determined
 from the proper-motion scatter of the seven points.

Absolute proper motions are obtained by applying the correction from Tab. 2
to each cluster as an offset. The error is derived from the quadrature sum of
the relative proper-motion error and that of the zero-point error.
We present in Table 4, the absolute proper motions and their errors, the 
Galactic coordinates, distances and radial velocities of the clusters.
The radial velocities are from H03. 
There are two distance values: the adopted distance is the one from H03, the
alternative one is a value given for clusters that may be on the 
far or near side of the bulge (where the Galactic center is taken to be at 
8.0 kpc from the Sun), according to likely distance uncertainties. 
For these clusters, the change in distance is such that they flip their location 
with respect to the Galactic center (from near to far and opposite).  This  
alters the velocity components in cylindrical coordinates $\Pi$ and $\Theta$
by more than their formal error. Essentially they can change sign, and/or
interchange their values.
One example of the uncertainty in distance is NGC 6287. While H03 provides a 
distance of 9.3 kpc, Lee et al. (2001) determine a distance of 7.4 kpc.
For NGC 6287, we will adopt the Lee et al. (2001)  distance estimate as 
an alternative value. For the rest of the
clusters, we simply adopt a value that will place them on the opposite 
side of the Galactic center than that given by the H03 distance, and is within
reasonable distance errors (or the order of $10\%$).

\section{Velocities}

Velocities are derived from the absolute proper motions, distances and radial velocities 
listed in Tab 4.
We assume that the Sun is at $(X, Y, Z) = 
(8.0,0.0,0.0)$ kpc,
the Local Standard of Rest (LSR) velocity is 
$(U, V, W) = (0,220,0)$ km s$^{-1}$, and the Sun peculiar motion is
$(U, V, W) = (-10.00,5.25,7.17)$ km s$^{-1}$ (Dehnen \& Binney 1998).
Velocity errors are determined from the formal errors in the proper motions, radial velocities
and a $10\%$ distance error. 

In Table 5 we list the velocity components in a Galactic restframe,
along with $(U,V,W)$ directions defined by the location of the LSR,
and along ($\Pi,\Theta$) directions in a cylindrical coordinate system.
$\Pi$ is along the radial direction from the Galactic center to the cluster location 
projected on to the Galactic plane, 
and positive outward from the Galactic center, while 
$\Theta$ is perpendicular to the radial direction and positive toward Galactic rotation. 
Together with the velocity components we  list
the cluster location in $(R_{GC},Z)$ coordinates, where 
$R_{GC} = \sqrt(X^2+Y^2)$,  the cluster metallicity and integrated
absolute magnitude ($M_V$) (H03). For clusters within 4 kpc from the Galactic 
center, we also provide the velocity of the bar at the location of the cluster,
assuming the bar rotates as a solid body with an angular velocity of
60 km s$^{-1}$ kpc$^{-1}$ (Gerhard 2002 and references therein). This velocity
is to be compared with the $\Theta$ component.
Table 6 is similar to Table 5, only for the subsample of clusters with
alternative distances.

\section{Discussion}

In Figure 8 we show the cluster locations and their velocities;
to the right it is shown a zoomed-in version of the left panels with 
clusters labeled for easier identification.  There are two projections 
in Fig 8: in the Galactic plane (top), and perpendicular to the Galactic
plane in the $(X,Z)$ plane (bottom). Metal rich ([Fe/H]$ > -1.0$) clusters are
shown with filled symbols, metal poor with open symbols.
The LSR is also shown with a square symbol together with its 
220 km s$^{-1}$ velocity vector.
The ellipse represents the bar which is drawn here at a $20\arcdeg$
orientation from the Sun-Galactic center direction, a semimajor axis
of 3.0 kpc, and an axis ratio of 10:3:3. These values are chosen from
the bar parameters summarized in Gerhard (2002), although he gives 
smaller values for the semimajor axis in agreement with 
studies based on scale lengths determinations from starcounts.
However, Bissantz et al. (2003) built a model of the inner Galaxy based on
COBE/DIRBE data and red clump stars, and constrain the corotation radius to be
$3.4\pm0.3$ kpc; this radius is known to constrain the maximum radius of the bar.
Babusiaux \& Gilmore (2005) determine
that the bar radius is between 2.5 and 3.0 kpc. 
Therefore, here we adopt a 3.0 kpc semimajor axis. 
The dotted lines indicate a 3.4-kpc circle representing the corotation radius,
and the solar circle radius of 8 kpc.   
In this discussion we do not consider the long bar, recently found by Benjamin et al. (2009),
which has a much thinner configuration than the traditional bar, i.e., a scale height
of 100 pc, while our clusters are at distances of 1 kpc or more from the plane.

In what follows we describe the halo as a population with means in 
each velocity component equal to 0 km  s$^{-1}$ and dispersions
$(\sigma_{\Pi},\sigma_{\Theta},\sigma_{W}) = (140, 90, 90)$ km s$^{-1}$ (e.g. Smith et al. 2009,
Carollo et al. 2010). For the thick disk we adopt the means in $\Pi$ and $W$ equal to 0 km s$^{-1}$ 
and $\Theta = 200 - 30 \times |z|$ km s$^{-1}$, where $|z|$ is the distance from the Galactic 
plane in kpc.
The thick disk dispersions adopted are  $(\sigma_{\Pi},\sigma_{\Theta},\sigma_{W}) = 
(80, 60, 40)$ km s$^{-1}$ (Girard et al. 2006, Carollo et al. 2010).

The two clusters located far from the bulge region, NGC 6284 and NGC 6356
both have velocities consistent with thick-disk membership: low
$\Pi$ and $W$ components and high $\Theta$. 
Based on kinematics alone, NGC 6356 has only a $50\%$ higher chance 
to belong to the thick disk than to the halo. However, taken together with the 
high metallicity, this cluster is better classified as a thick-disk cluster.

While NGC 6356 is a metal 
rich cluster, and therefore expected to belong to the thick disk, 
NGC 6284 is a mildly metal poor cluster. We therefore add one more cluster to
the sample of metal-poor ([Fe/H] $ < -1.0$) clusters with thick-disk 
like orbits: NGC 6626 (M 28) - Cudworth \& Hanson 1993, 
NGC 6752 - Paper I, NGC 6266 (M 62) - Paper IV, and 
possibly NGC 6254 (M10) and NGC 6171 - Paper III.

For all clusters within $R_{GC} \sim 1$ kpc, the distance ambiguity
makes their velocity interpretation somewhat difficult. 
It appears safe to say that two metal poor clusters, NGC 6273 and NGC 6293 
have velocities consistent with halo membership under both distance
estimates (Tab. 5 and 6) due to their high $W$ velocity components.
NGC 6342, a metal rich cluster, has velocities consistent with the thick disk or 
a ``nominal'' bulge described as a spheroid with a rotation 
velocity of $63 \pm 41$ km s$^{-1}$ and a dispersion of $110$ km s$^{-1}$ (C\^{o}t\'{e} 1999, see 
also Section 5.1).

NGC 6287, another metal poor cluster has velocity components that 
argue in favor of membership to the bar (Tab. 5 and 6). However 
its location 1.4 kpc (Tab. 6) above the Galactic plane 
makes this association less likely. Therefore, we associate this cluster with the halo.
The last inner 2-kpc, metal-poor cluster, NGC 6333 has velocities consistent with
thick disk membership; however, the distance uncertainty, makes this membership
less evident, with a non-negligible chance to belong to the halo.

\subsection{The Case of NGC 6388 and NGC 6441}

Clusters NGC 6388 and NGC 6441 are known to have bimodal horizontal branches as well as
very extended blue horizontal branches (EHB) in spite of their high,
thick-disk like  metallicity (Rich et al. 1997,
Piotto et al. 2002, Busso et al. 2007). They are also very massive: the third and fifth in rank after 
clusters $\omega$ Cen, M 54 (at the center of Sagittarius dwarf galaxy) and NGC 2419 (H03, Yoon et al. 2008).
Thus their properties fit the picture proposed by
Lee et al. (2007) where EHB clusters are predominantly massive clusters with double or multiple 
populations, of which at least one is enriched in He.
Recently, HST photometry has
unambiguously shown that the EHB clusters observed have multiple main sequences
and subgiant branches (e.g., Piotto 2008 and references therein), and there is a growing body 
of evidence that He enrichment is responsible for the extended blue part of the 
horizontal branch as well as for the blue main sequences (Piotto 2008).
For NGC 6388, Piotto (2008) shows that it has a double subgiant branch, but there is no clear 
evidence that NGC 6388 and 6441 have populations with multiple metallicities as is the case of
$\omega$ Cen. At any rate, the existence of a He-rich population is best explained by 
self enrichment of the system from a previous generation of 5 to 10 M$_{\odot}$ stars 
(Renzini 2008 and references therein).
As discussed by Renzini (2008), the major problem with self 
enrichment is that
the first-generation population currently seen in the cluster can not produce sufficient 
He to account for the existing He-rich, second-generation population seen in the same cluster.
Thus one has to invoke additional material brought in from an originally much more massive system
than the present-day cluster (e.g. Bekki \& Norris 2006). Also, in the case of $\omega$ Cen, 
Gnedin et al. (2002) show that the self-enrichment scenario is at odds with the present-day orbit
that implies many disk crossings that would strip the cluster of its gas produced, for 
instance, by AGB stars.
This may also be the case for NGC 6338 and NGC 6441, with their low energy, in plane orbits
as suggested by their velocities.
The above arguments naturally imply the existence of these type of clusters at the bottom of a 
deep potential well, a system resembling more a dwarf galaxy, rather than a cluster. 
Lee et al. (2007) as well as Bekki \& Norris (2006) proposed that these EHB clusters 
are of extragalactic origin, formed probably  
as nuclei in dwarf satellites that were accreted by the Milky Way, with
$\omega$ Cen being the prototype.

The kinematical evidence from the present work clearly shows that these two clusters are 
not members of the thick disk despite their high metallicity. 
Also, the low velocity components indicate that NGC 6388 and NGC 6441 are at or near 
apocenter. Thus they are confined to 
the inner $\sim 4$ kpc of the Galaxy. However, their velocity components 
and for NGC 6388, its location too, clearly 
show that they do not belong to the bar (Fig. 7).

As such, they can be classified
as members of a kinematically hotter system than the thick disk, a ``nominal'' bulge
with a spherical distribution that peaks at the Galactic center. 
Metal-rich globular clusters within $\sim 4$ kpc from the Galactic center
appear to be consistent with such a picture (Minniti 1995,
 C\^{o}t\'{e} 1999, Bica et al. 2006), although some metal-rich globular clusters ---
those of low luminosity --- may
belong to the bar (C\^{o}t\'{e} 1999, Burkert \& Smith 1997).
The existence of such a system distinct from the well-known bar 
mentioned above is still a matter of debate (see e.g. Howard et al. 2009). 
Nevertheless, assuming such a system exists, we adopt
its kinematic properties from C\^{o}t\'{e} (1999) who, based on radial velocities,
distance and geometry, determines a mean rotation and velocity dispersion 
for a sample of 19 metal rich 
clusters within 4 kpc from the Galactic center.
He obtains a rotation velocity of $63 \pm 41$ km s$^{-1}$ and a dispersion of
$110$ km s$^{-1}$. Using our $\Theta$ components and their errors, and the
mean rotation and the dispersion from C\^{o}t\'{e} (1999), we obtain that 
both NGC 6388 and NGC 6441 are consistent with membership in 
this system (within $0.9\sigma$ and $0.4\sigma$ from the mean respectively).

Their very similar velocity components 
and similar  Galactic locations imply similar orbits.
Motivated by this similarity, as well as by the similarity 
of their peculiar physical properties which here we parametrize only by
the integrated absolute magnitude and metallicity, we perform a statistical experiment
to quantify the chance of spatial and velocity coincidence.
First, we choose a representative Galactocentric radius interval for the two clusters.
Under the assumption that the error in the distance is $10\%$ of the distance,
NGC 6388 is located within $3.2\pm0.7$ kpc from the Galactic center, while 
NGC 6441 is  within $3.9\pm 1.0$ kpc. Thus the representative Galactocentric 
radius interval is chosen to be from 2.5 to 5.0 kpc. In this interval there are only 5 globular 
clusters with [Fe/H] $> -1.0$, and $M_V < -8.0$ (H03). The magnitude limit is
taken from the analysis of Lee et al. (2007), which shows that the majority of EHB 
clusters are brighter than this limit. We also assume that bright, massive clusters are 
not missing from our sample due to foreground extinction.
We will therefore look for the 
spatial and velocity separation for a set of five clusters in 
a distribution that resembles the metal-rich globular-cluster distribution in the
inner Galaxy, and compare with the same quantities for NGC 6388 and NGC 6441.
The spatial separation between NGC 6388 and NGC 6441 is $2.3 \pm 1.3$ kpc,
while the velocity separation is $83 \pm 38$ km s$^{-1}$.
We generate a spatial distribution that obeys a density law of the
form $\rho(r) \propto r^{-3.2}$ (Bica et al. 2006), and is located between 2.5 and 5.0 kpc. 
To each point, velocity components 
are assigned from Gaussian velocity distributions with means
$(\Pi, \Theta, W) = (0, 63, 0)$ km s$^{-1}$ and dispersions 
$(\sigma_{\Pi}, \sigma_{\Theta}, \sigma_W) = (100, 100, 100)$ km s$^{-1}$ that resembles 
the C\^{o}t\'{e} (1999) description of bulge clusters.
From this population we extract 5 points and calculate their spatial and
velocity separations. We retain only the smallest spatial separation and its
corresponding velocity separation. Then we repeat this experiment 20000 times.
We obtain that 
there is a $3\%$ chance to obtain lower spatial and velocity separations 
than those of the two clusters. If we consider a $1\sigma$ uncertainty in the
clusters' velocity separation, then this chance is $8\%$.
While this is a non-negligible chance coincidence,
it is however rather low, and makes the issue of a common dynamical origin of the 
two clusters an intriguing possibility.


\section{Conclusions}

Using SPM photographic and CCD data, we determine the absolute proper motions of nine 
globular clusters: five metal poor ([Fe/H] $< -1.0$) and four metal rich clusters.
Seven clusters are located within $\sim 4$ kpc from the Galactic center, 
and two are located at a distance of $ \sim 7.5$ kpc from the Galactic center,
in the background of the bulge.

We determine 3D velocities by combining our proper motion data with
radial velocities and distances from the literature. 
From the five metal poor clusters, three clusters --- NGC 6273, NGC 6287 and NGC 6293 ---
appear to belong to a kinematically hot
system, the inner halo. From the remaining two metal poor clusters, NGC 6284 belongs to the thick 
disk, while NGC 6333 has an ambiguous membership (between halo and thick disk) due to
the distance uncertainty.

From the metal rich clusters, NGC 6388 and NGC 6441 have kinematics clearly inconsistent
with thick disk membership; they can be thought of as members of
a ``nominal'' bulge (not the bar), i.e., a spheroid with low rotation and high velocity dispersion.
More interestingly, the two clusters have very similar velocity components. This taken
together with their close spatial location and peculiar but similar 
physical properties argue in favor of a common dynamical origin of the two clusters.
The remaining two metal rich clusters NGC 6356 and NGC 6342 have kinematics consistent
with thick disk membership, but velocity uncertainties stemming from distance
and proper-motion uncertainties do not allow us to
rule out their membership to a hotter system. While NGC 6342 is within 2 kpc from the 
Galactic center, and thus can be part of the bulge,
NGC 6356 is located outside the bulge region. Thus its location, metalicity and velocities combined
argue in favor of thick-disk membership.

\acknowledgments

This work is supported by NFS grants AST 04-07292, AST 04-07293 and 
AST 09-08996. We wish to thank Ralph Launhardt for providing the 
data for the bulge map in Fig. 4.
We are grateful to David Herrera, Kathy Vieira and Young Sun Lee for their help with the CCD reductions. 

This publication makes use of data products from the Two Micron All Sky Survey, which is a joint project of the University of Massachusetts and the Infrared
Processing and Analysis Center/California Institute of Technology, funded 
by  NASA and  NSF.

\clearpage

\begin{table}[htb]
\begin{center}
\caption{Field Centers and Observational Material}
\begin{tabular}{crrrrllc}
\tableline
\\
\multicolumn{1}{c}{SPM} & 
\multicolumn{1}{c}{R.A.} &
\multicolumn{1}{c}{Dec.} &
\multicolumn{1}{c}{l} &
\multicolumn{1}{c}{b} &
\multicolumn{1}{l}{Photographic Plates} &
\multicolumn{1}{l}{SPM CCD$^{a}$} &
\multicolumn{1}{c}{Other$^{b}$} \\
& \multicolumn{2}{c}{(B1950)} & \multicolumn{2}{c}{($\arcdeg$)} & \multicolumn{1}{l}{Pairs~~~~~~Epoch} &
\multicolumn{1}{l}{$N_{S}~N_{C}~~~~~~~$Epoch} & 
\multicolumn{1}{c}{(\% area)} \\
\tableline
\\
578 & 255 & -25 & 357.9 & 10.0 & 2(BY) 1967.45, 1969.36 & $~71~23~~2003-2004$ & 23 \\
651 & 260 & -20 & 4.7 & 9.1 & 1(BY) 1973.50 & $~96~~9~~2004$ & 1 \\
307 & 264 & -45 & 345.6 & -7.4 & 2(BY) 1968.41, 1969.63 & $106~~6~~2004-2005, 2007$ & 4 \\
436 & 265 & -35 & 354.6 & -2.8 & 2(BY) 1967.36, 1989.42 & $~~~0~~6~~2007$ & 0 \\
\tableline 
\multicolumn{8}{l}{{$^a~$}Number of frames taken by the main survey ($N_S$) and on clusters ($N_C$).} \\
\multicolumn{8}{l}{{$^b~$}2nd epoch positions from HIPPARCOS, TYCHO2, UCAC2, or 2MASS.}
\end{tabular}
\end{center}
\end{table}

\begin{table}[htb]
\begin{center}
\caption{Corrections to Absolute Proper Motions}
\begin{tabular}{rrrrl}
\tableline
\multicolumn{1}{c}{SPM} &
\multicolumn{1}{c}{$\Delta\mu_{\alpha}^*$ (H-SPM)} &
\multicolumn{1}{c}{$\Delta\mu_{\delta}$ (H-SPM)} &
\multicolumn{1}{c}{N$^c$} & \multicolumn{1}{c}{Clusters} \\
 & \multicolumn{1}{c}{(mas~yr$^{-1}$)} &
 \multicolumn{1}{c}{(mas~yr$^{-1}$)} & \\
\tableline
\\
578 & -2.54(0.33) & -4.95(0.36)& 92 & NGC 6273, 6284, 6287, 6293\\
651 & -1.46(0.43) & -4.26(0.33)& 84 & NGC 6333, 6342, 6356 \\
307 & -1.08(0.30) & -5.02(0.38)& 152 & NGC 6388 \\
436 & -1.07(0.25) & -3.04(0.40)& 56 - 101 & NGC 6441 \\
\tableline
\multicolumn{5}{l}{{$^c~$} for SPM field 436, we show the range of
 $Hipparcos$ stars used for all image orders } \\
\multicolumn{5}{l}{and plate-pair solutions.} 
\end{tabular}
\end{center}
\end{table}

\begin{table}[htb]
\begin{center}
\caption{Relative Proper Motion of the Clusters}
\begin{tabular}{lrrrrr}
\tableline
\multicolumn{1}{c}{NGC} & \multicolumn{1}{c}{$\mu_{\alpha}^*$} & \multicolumn{1}{c}{$\mu_{\delta}$} &
\multicolumn{1}{c}{$K_{faint}$} & \multicolumn{1}{c}{$N_{sample}$} & \multicolumn{1}{c}{Method} \\
 & \multicolumn{1}{c}{(mas~yr$^{-1}$)} & \multicolumn{1}{c}{(mas~yr$^{-1}$)} & & & \\
\tableline
 6273/M 19 & -0.33(0.36)& 4.50(0.36) & 13.0 & 301 & 2G fit \\
 6284 & -1.13(0.55) & -0.44(0.75) & 12.0& 39 & pplot \\
 6287 & -1.15(0.82) & 1.41(0.59) & 12.0 & 52 & pplot \\
 6293 & 2.78(0.78) & -0.19(0.61) & 12.0 & 49 & pplot \\
 6333/M 9 & 0.89(0.37) & 0.56(0.37) & 13.5& 201 & 2G fit \\
 6342 & -1.31(0.56) & -1.88(0.56) & 13.7& 296 & 2G fit \\
 6356 & -1.68(0.53) & 0.61(0.41) & 13.2 & 97 & pplot \\
 6388 & 0.18(0.34) & 1.59(0.34) & 13.0 & 240 & 2G fit\\
 6441 & -1.79(0.37)& -0.41(0.65) & 12.0 & 7 & ave \\
\tableline
\end{tabular}
\end{center}
\end{table}

\begin{table}
\begin{center}
\caption{Absolute Proper Motions, Galactic Coordinates, Distance and Radial Velocity}
\begin{tabular}{rrrrrrrr}
\tableline
 \multicolumn{1}{c}{NGC} &   \multicolumn{1}{c}{$\mu_{\alpha}^*$} & \multicolumn{1}{c}{$\mu_{\delta}$} & 
\multicolumn{1}{c}{$l$} & \multicolumn{1}{c}{$b$} & 
\multicolumn{1}{c}{$d_{adopt}$} & \multicolumn{1}{c}{$d_{alt}$} &
\multicolumn{1}{c}{$V_{rad}$} \\
 & \multicolumn{1}{c}{(mas~yr$^{-1}$)} & \multicolumn{1}{c}{(mas~yr$^{-1}$)} & & & (kpc) & (kpc) & (km s$^{-1}$) \\
\tableline
 6273 & -2.86(0.49) & -0.45(0.51) & -3.1 & 9.4 & 8.6& 7.6 &135.0(4.0)\\ 
 6284 & -3.66(0.64) & -5.39(0.83) & -1.7 & 9.9 & 15.3 & ... & 27.6(1.7) \\ 
 6287 & -3.68(0.88) & -3.54(0.69) & 0.1 & 11.0 & 9.3 & 7.4 & -288.8(3.5)  \\ 
 6293 & 0.26(0.85) & -5.14(0.71) &  -2.4 & 7.8 & 8.8 & 7.8 & -146.2(1.7) \\ 
 6333 & -0.57(0.57)& -3.70(0.50) & 5.5 & 10.7 & 7.9 & 8.9 & 229.1(7.0) \\ 
 6342 & -2.77(0.71)& -5.84(0.65) & 4.9 & 9.7 & 8.6 & 7.6 & 116.2(1.6) \\ 
 6356 & -3.14(0.68) & -3.65(0.53) & 6.7 & 10.2 & 15.2 & ... & 27.0(4.3)  \\ 
 6388 & -1.90(0.45) & -3.83(0.51)  & -14.4 & -6.7 & 10.0 & ... & 81.2(1.2) \\ 
 6441 & -2.86(0.45) & -3.45(0.76) &   -6.5 & -5.0 & 11.7 & ... & 16.4(1.2)  \\ 
\tableline
\end{tabular}
\end{center}
\end{table}

\begin{table}
\begin{center}
\caption{Positions and Velocities for the Adopted Distance}
\begin{tabular}{rrrrrrrrrrr}
\tableline
\multicolumn{1}{c}{NGC} & \multicolumn{1}{c}{[Fe/H]} &
\multicolumn{1}{c}{$M_V$} & \multicolumn{1}{c}{$Z$} &\multicolumn{1}{c}{$R_{GC}$} &
\multicolumn{1}{c}{$U$} & \multicolumn{1}{c}{$V$} & \multicolumn{1}{c}{$W$} & \multicolumn{1}{c}{$\Pi$} &\multicolumn{1}{c}{$\Theta$} & \multicolumn{1}{c}{$\Omega_{b} \times R_{GC}$} \\
 & & & \multicolumn{2}{c}{(kpc)}  &
\multicolumn{3}{c}{(km s$^{-1}$)}   &
\multicolumn{3}{c}{(km s$^{-1}$)}   \\
\tableline
 6273 &-1.68 & -9.18 & 1.4 & 0.7 & -125(06)& 134(22)  & 110(22) & -5(16)& -183(16) & 42 \\
 6284 &-1.32 & -7.97 & 2.6 & 7.1 & -27(09) & -248(72) & -7(51) & 42(10) & 245(72) & ... \\
 6287 &-2.05 & -7.36 & 1.8   & 1.1 & 280(08)   & 3(40) & -10(36)& -280(08) & -8(40) & 66 \\
 6293 &-1.92 & -7.77 & 1.2 & 0.8 & 123(05) & 64(36) & -146(36) & -139(17) & -1(32) & 48 \\
 6333 &-1.75 & -7.94 & 1.5   & 0.8 & -257(08)  & 122(23)  & -9(21) & 26(22) & 283(11) & 48 \\
 6342 &-0.65 & -6.44 & 1.5  & 0.8 & -153(06) & -24(39)& -11(28)& 60(33) & 143(21) & 48 \\
 6356 &-0.50 & -8.52 & 2.7  & 7.1 & -69(11)  & -114(53) & 56(46) & 38(17) & 128(52) & ... \\
 6388 &-0.60 & -9.42 & -1.2 & 3.0 &-35(08) & 11(30) & -23(22) & 11(26) & -35(18) & 180 \\
 6441 &-0.53 & -9.64 & -1.0 & 3.8 & -2(06) & -21(46) & 44(31) & 9(17)& 19(43) & 228 \\
\tableline
\end{tabular}
\end{center}
\end{table}

\begin{table}
\begin{center}
\caption{Positions and Velocities for the Alternative Distance}
\begin{tabular}{rrrrrrrrrrr}
\tableline
\multicolumn{1}{c}{NGC} & \multicolumn{1}{c}{[Fe/H]} &
 \multicolumn{1}{c}{$M_V$} & \multicolumn{1}{c}{$Z$} & \multicolumn{1}{c}{$R_{GC}$} &
\multicolumn{1}{c}{$U$} & \multicolumn{1}{c}{$V$} & \multicolumn{1}{c}{$W$} & \multicolumn{1}{c}{$\Pi$} &\multicolumn{1}{c}{$\Theta$} & \multicolumn{1}{c}{$\Omega_{b} \times R_{GC}$} \\
 & & & \multicolumn{2}{c}{(kpc)}  &
\multicolumn{3}{c}{(km s$^{-1}$)} &
\multicolumn{3}{c}{(km s$^{-1}$)} \\
\tableline
 6273 &-1.68 & -9.18 &  1.2 & 0.7 & -127(05) & 144(20)  & 101(19) & -189(13)& 33(16) & 42 \\
 6287 &-1.32 & -7.36 &  1.4 & 0.7 & 279(07)   & 48(32) & -18(28)& 280(07) & 42(32) & 42 \\
 6293 &-1.92 & -7.77 &  1.1 & 0.4 & 125(05)  & 83(32) & -131(32) & 20(24) & 148(21) & 24 \\
 6333 &-1.75 & -7.94 &  1.7 & 1.1 & -260(09) & 106(26)  & -16(24) & 248(21) & 132(18) & 66 \\
 6342 &-0.65 & -6.44 &  1.3 & 0.8 &  -150(06)  & 6(34)& -7(25)& -92(26) & 118(22) & 48 \\
\tableline
\end{tabular}
\end{center}
\end{table}

\clearpage

\begin{figure}
\includegraphics[scale=0.7]{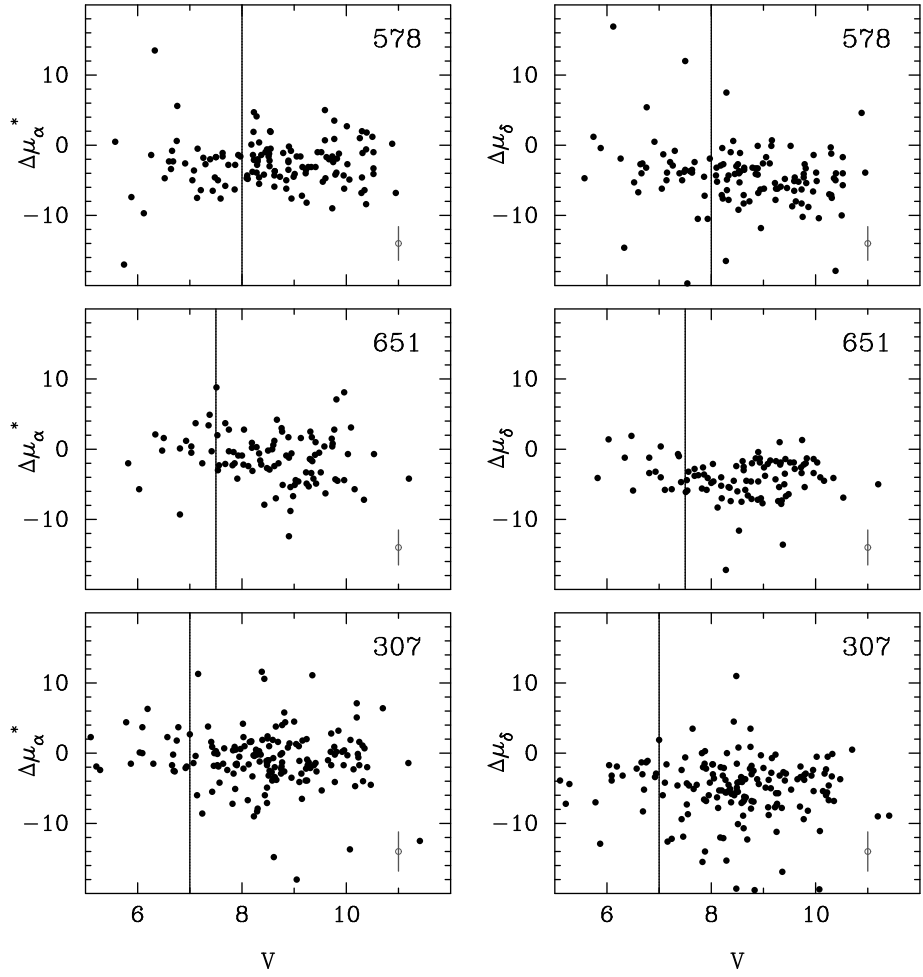}
\caption{Proper-motion differences (Hipparcos-SPM) as a function of magnitude for three fields. 
The vertical bar indicates the bright magnitude limit 
used in determining 
the mean corrections to absolute proper motion per field. 
A typical error bar is shown in the lower right.
The labels indicate the SPM field number. Proper-motion units here and throughout the paper are mas yr$^{-1}$.}
\end{figure}

\begin{figure}
\includegraphics[angle=-90,scale=0.7]{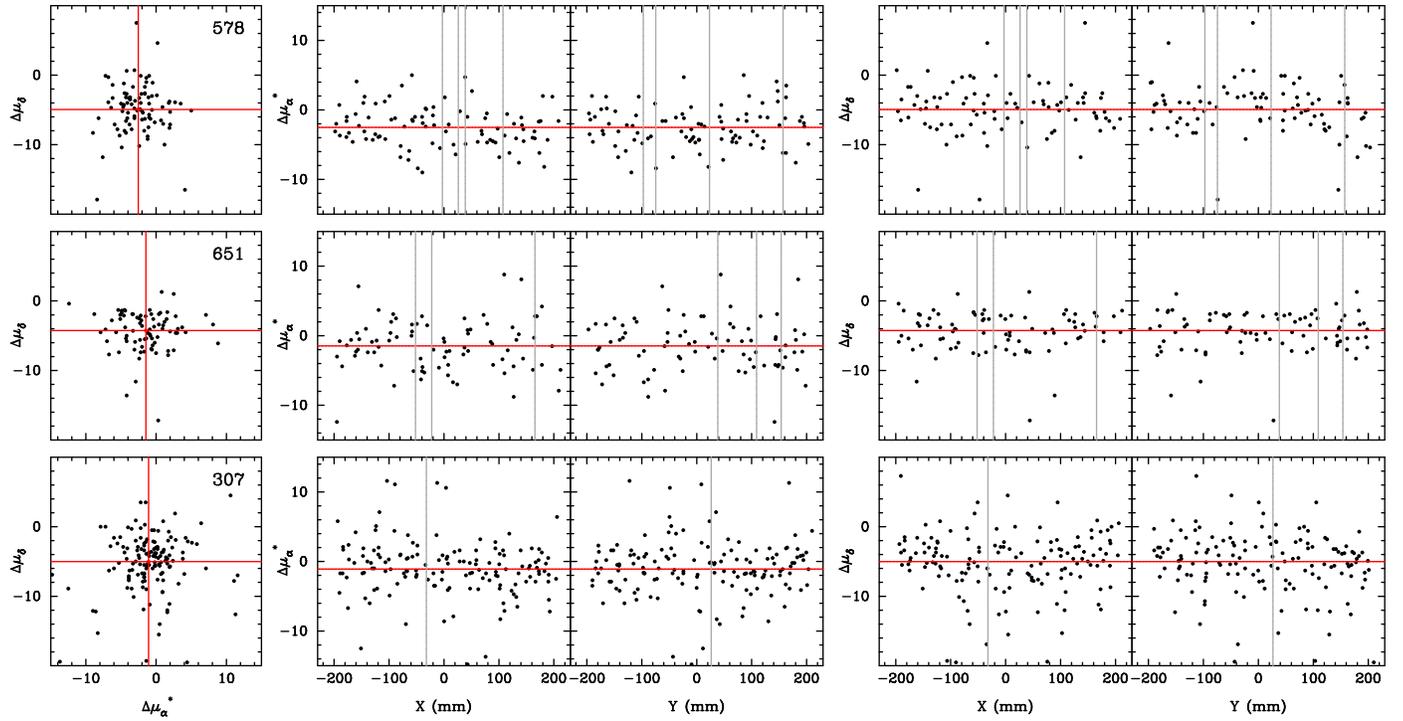}
\caption{Proper-motion differences (Hipparcos-SPM) in both coordinates (left panels)
and as a function of position on the plate (middle and right) for three fields. The red 
lines indicate the mean correction to absolute as determined in the text. The vertical lines show the position of the clusters on each field.}
\end{figure}

\begin{figure}
\includegraphics[scale=1.0]{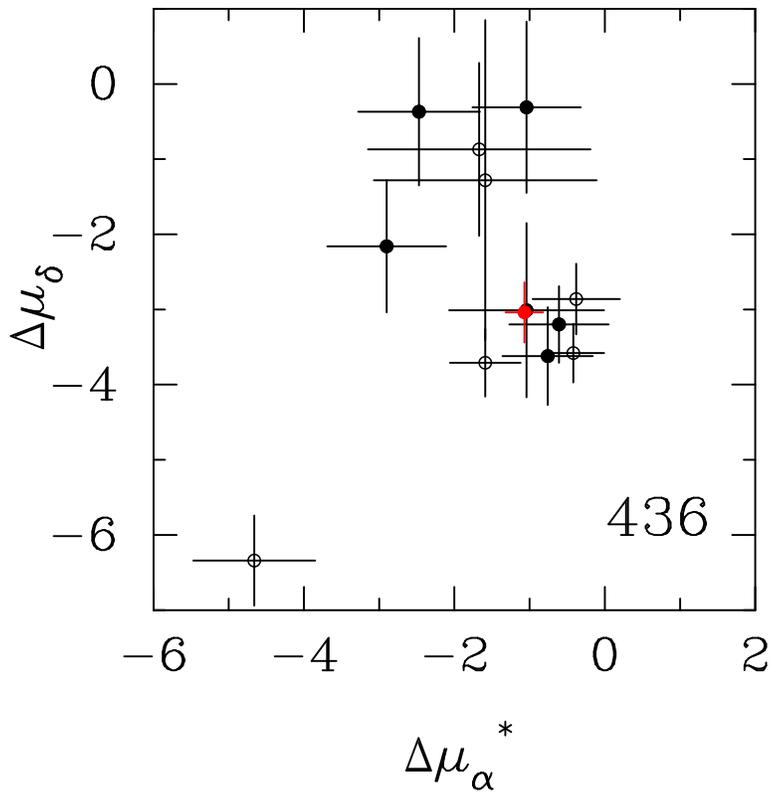}
\caption{Proper-motion differences (H-SPM) for various plate-pair and image-order solutions for
SPM field $\#436$. Open symbols show the visual plate solution, filled symbols, the blue plate solution. In red is shown the mean valued adopted (see text).}
\end{figure}

\begin{figure}
\includegraphics[scale=0.7]{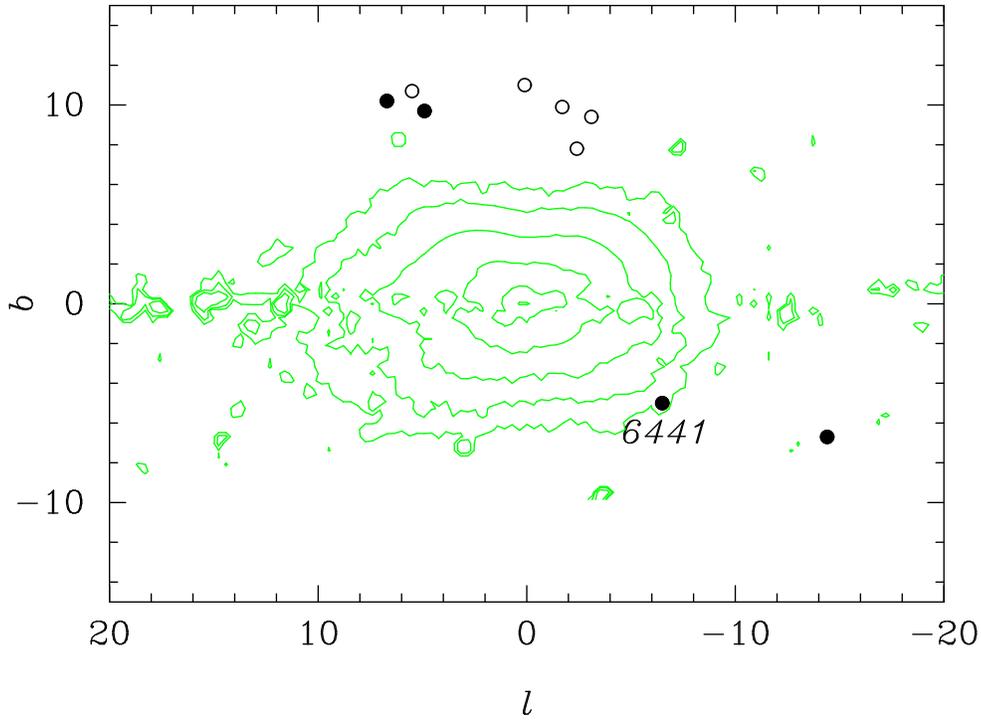}
\caption{The 2.2 $\mu$m surface brightness distribution map from Launhardt et al. (2002) (green contours)  and the globular clusters from this study.
Metal rich clusters ([Fe/H] $> -1.0$) are represented with filled circles,
metal poor with open circles. Note the location of NGC 6441 near the far end 
of the bar.}
\end{figure}

\begin{figure}
\includegraphics[angle=-90,scale=0.7]{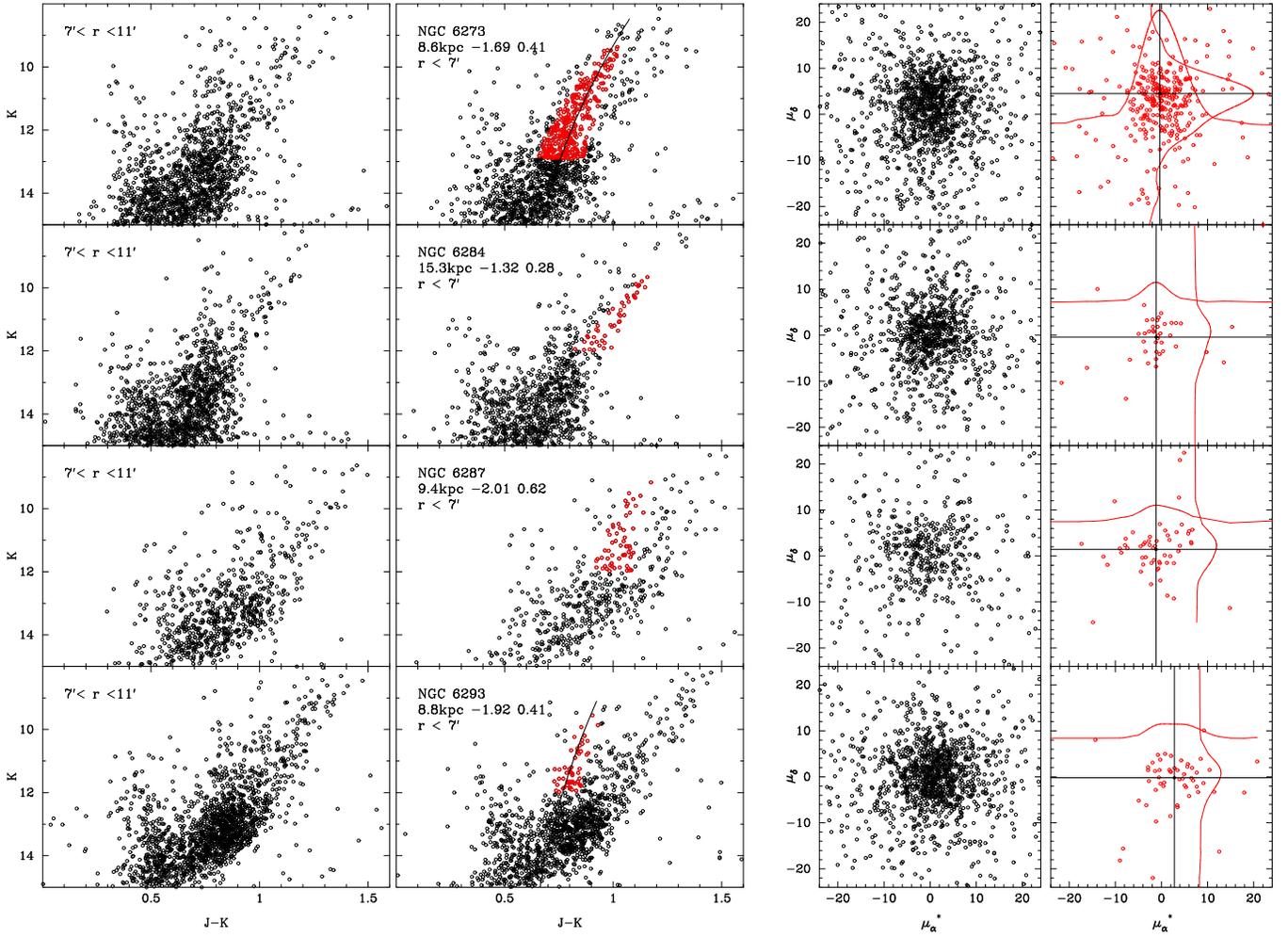}
\caption{2MASS CMDs and VPDs of the relative proper motions
 for four clusters: NGC 6273, NGC 6284, NGC 6287, and NGC 6293. The first column  from the left shows CMDs
for stars in a ring around the cluster, the second column shows the cluster region.
In each panel we specify the radius from the cluster center covered by the sample; for the cluster region, we also list the 
distance to the cluster, the metallicity and the reddening $E_{B-V}$. Cluster stars selected for the proper-motion determination 
are selected along the giant branch, and are shown in red. For some clusters, we also show the ridge line of the giant branch as 
given by Valenti et al. (2004). The third column shows the VPD for the entire cluster region ($r < 7\arcmin$), while the fourth 
column shows the VPD of the cluster sample corresponding to the giant branch in the second column. 
Proper-motion distributions are also shown, and the adopted mean (see text) is indicated with the black lines.}
\end{figure}

\begin{figure}
\includegraphics[angle=-90,scale=0.7]{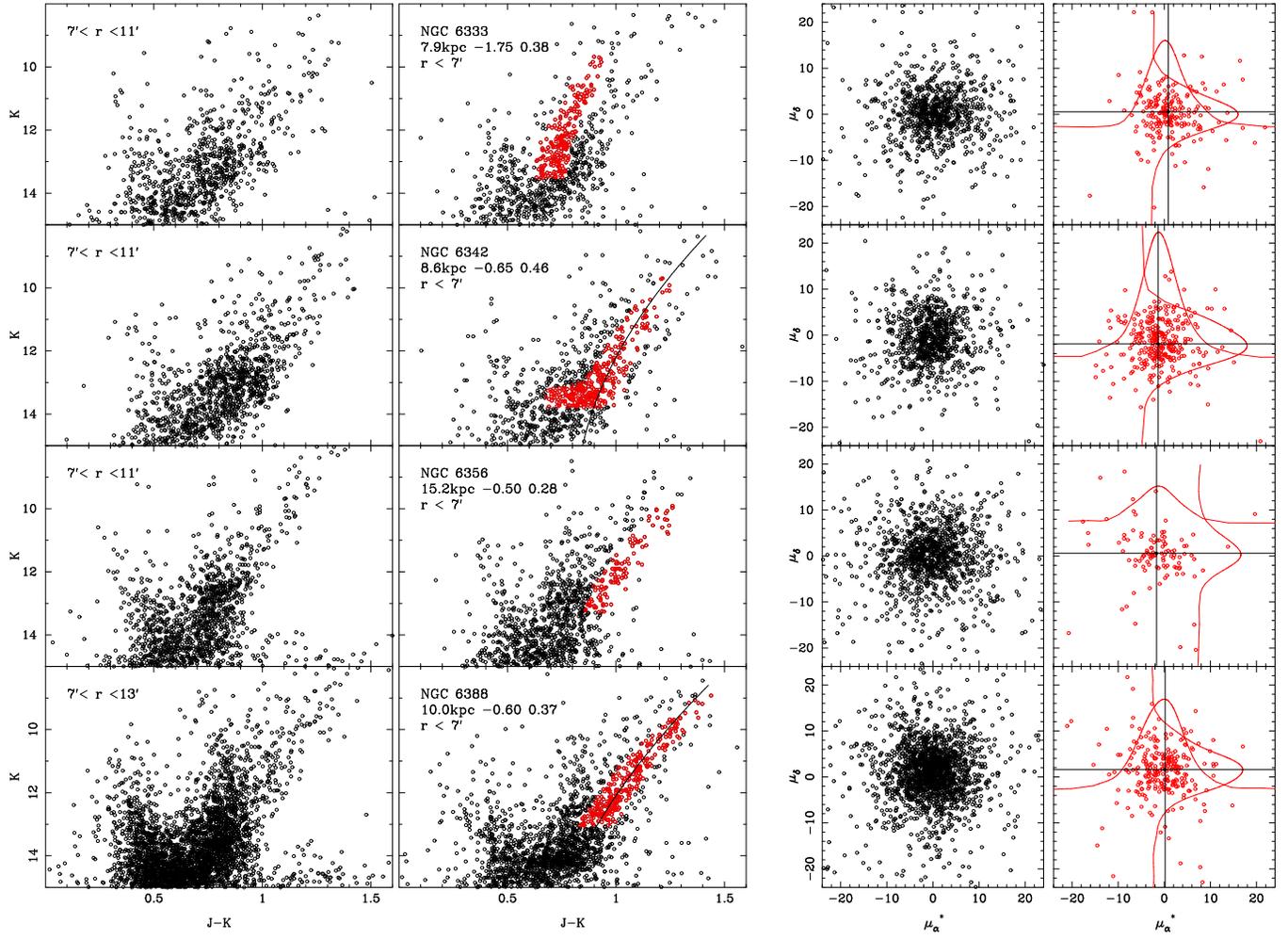}
\caption{Same as Fig. 4, only for NGC 6333, NGC 6342, NGC 6356 and NGC 6388.}
\end{figure}

\begin{figure}
\includegraphics[angle=-90,scale=0.7]{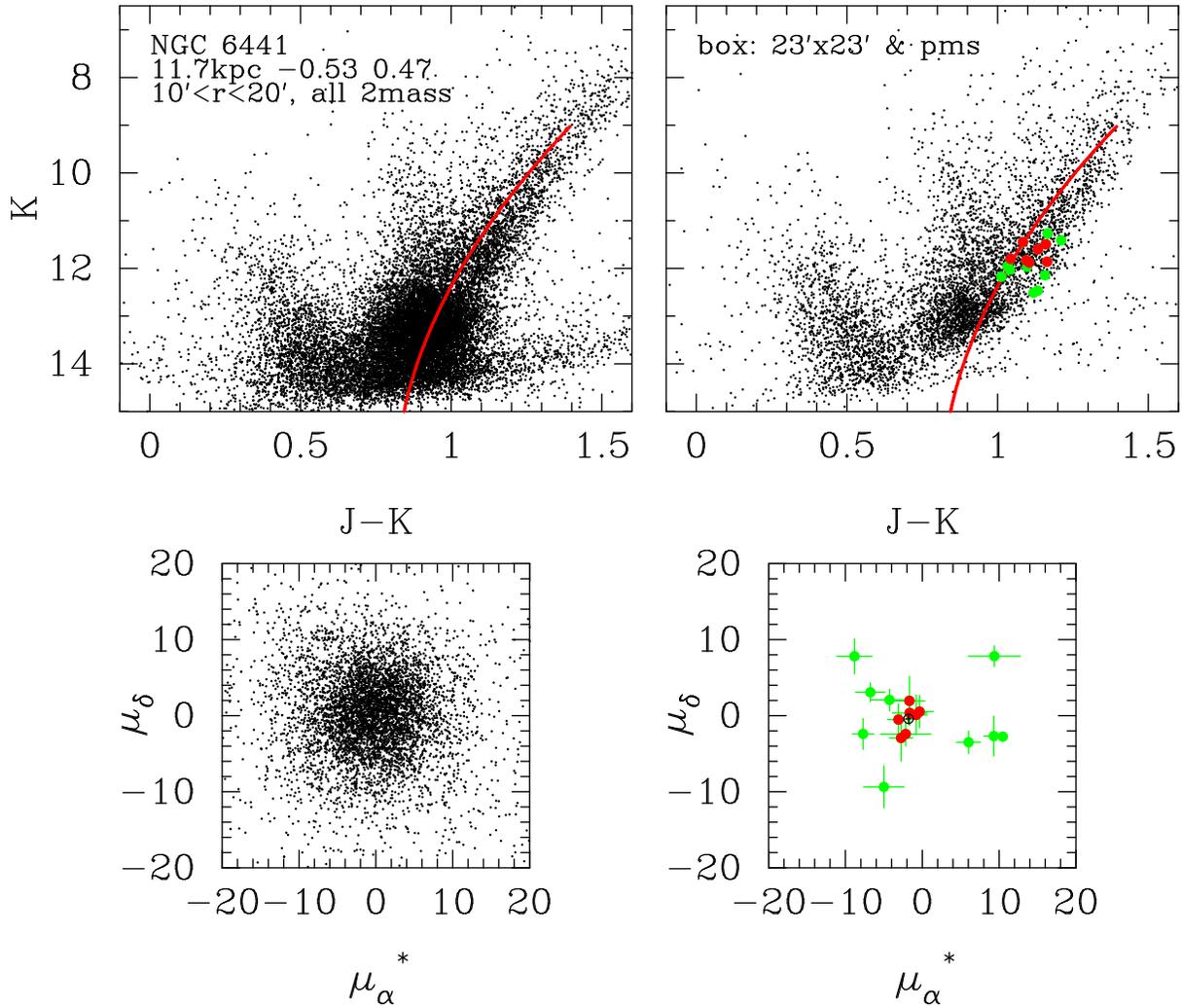}
\caption{2MASS CMD for a region $10\arcmin <r< 20\arcmin$ centered on NGC 6441 (top left).
The  ridge line from Valenti et al. (2004) is shown in red: clearly it is offset from the 
prominent giant branch seen in the CMD. A smaller region centered on the cluster for which proper motions are derived
is shown in the top-right panel. The filled symbols show the
radial-velocity members from Gratton et al. (2007). In red are shown a subsample used for the mean proper motion of the cluster. The bottom left panel shows
 the relative proper motions for the entire sample from top-right panel. The 
bottom-right panel shows the proper-motions for the radial-velocity selected sample, with red symbols for the stars used in the determination of the mean. 
The black open circle indicates the mean adopted; its error bars are of the size of the symbol.}
\end{figure}

\begin{figure}
\includegraphics[angle=-90,scale=0.80]{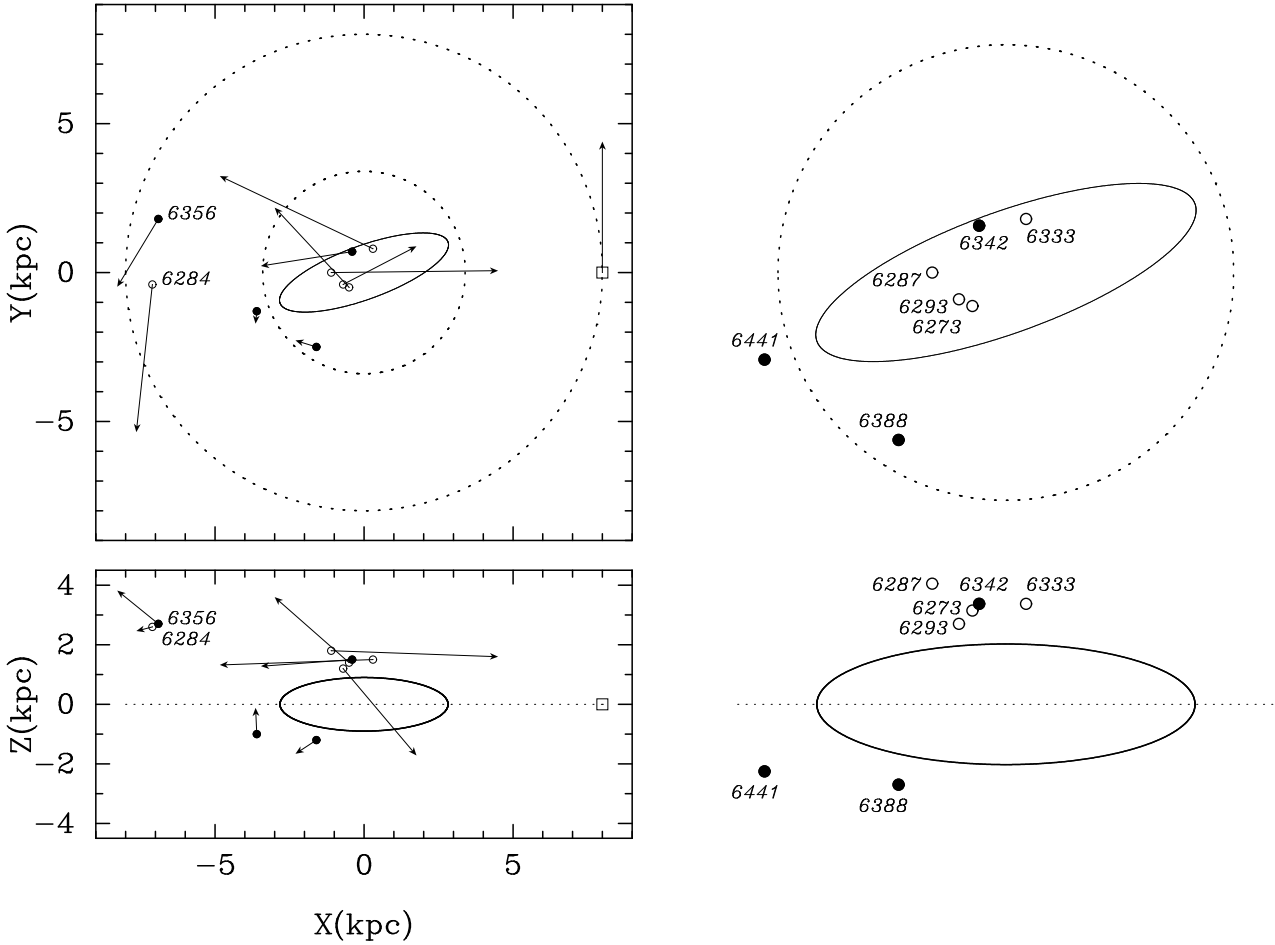}
\caption{Positions and $UVW$-velocities of the clusters 
projected onto the Galactic plane (top) and 
perpendicular to the Galactic plane (bottom). Filled symbols represent the
metal rich ([Fe/H] $\ge$ -1.0) clusters; open, the metal poor clusters.
The local standard of rest (and its 220 km s$^{-1}$ velocity) is shown with a square. The ellipse represents the Galactic bar. The dotted circles, mark the bar corotation radius, and the solar circle. To the right is shown a zoomed-in version of the left panels, with clusters labeled.}
\end{figure}

\end{document}